\documentclass[aps,pra,reprint,twocolumn,superscriptaddress,showpacs,showkeys,superscriptaddress,longbibliography]{revtex4-1}
\usepackage{amsmath,amssymb,amstext}
\usepackage[usenames,dvipsnames]{color}
\usepackage{graphicx}
\usepackage{bm,bbold,braket}
\usepackage{natbib}
\usepackage{txfonts, comment,stmaryrd}
\usepackage{dcolumn}
\usepackage{color}
\usepackage{xcolor}
\colorlet{rn}{red}
\colorlet{an}{blue}
\usepackage[english]{babel}
\usepackage{mathcomp}
\usepackage{wasysym}
\usepackage[colorlinks,bookmarks=false,citecolor=blue,linkcolor=red,urlcolor=blue]{hyperref}
\begin{document}

\title{Maximally entangled Rydberg-atom pairs via Landau-Zener sweeps}
\author{Dhiya Varghese}
\affiliation{Indian Institute of Science Education and Research, Pune- 411008, India}
\author{Sebastian W\"uster}
\affiliation{Indian Institute of Science Education and Research, Bhopal, India}
\author{Weibin Li}
\affiliation{School of Physics and Astronomy, University of Nottingham, NG7 2R8, United Kingdom}
\author{Rejish Nath} 
\affiliation{Indian Institute of Science Education and Research, Pune- 411008, India}
%\date{\today}

\begin{abstract}
We analyze the formation of maximally entangled Rydberg atom pairs subjected to Landau-Zener sweeps of the atom-light detuning. Though the populations reach a steady value at longer times, the phases evolve continuously, leading to periodic oscillations in the entanglement entropy. The local unitary equivalence between the obtained maximally entangled states and the Bell states is verified by computing the polynomial invariants. Finally, we study the effect of spontaneous emission from the Rydberg state of rubidium atoms on the correlation dynamics and show that the oscillatory dynamics persists for high-lying Rydberg states. Our study may offer novel ways to generate maximally entangled states, quantum gates and exotic quantum matter in arrays of Rydberg atoms through Landau Zener sweeps.
\end{abstract}



\maketitle
%\tableofcontents

\section{Introduction}
%
%% entanglement and bell pair
Entanglement is an essential resource in quantum technology \cite{eke91,ben92,div00,joz03, tot14}. Controlled unitary processes/quantum gates \cite{zhe00, mol08, wil10, lad10,zha10, lin16, zen17}, dissipative state engineering \cite{ple99,kra08,kas11,col22} or a combination of both \cite{lin13} are used to generate entanglement between two qubits. Maximally entangled qubit pairs including the Bell states are realized in different physical platforms such as ions \cite{tur98,lin13, lin16,col22}, atom-photon hybrid system \cite{bli04}, superconducting qubits \cite{dic09,sha13,kim16,liu16},  Josephson phase qubits \cite{ans09}, and Rydberg atoms \cite{hag97, jau16}. The latter are at the forefront of studies in quantum information processing and many-body quantum simulations because of the prodigious Rydberg-Rydberg interactions (RRIs) and the versatility in engineering them \cite{saf10, saf16, bro20,mor21}.

%% LZ model
A Landau-Zener (LZ) transition between two energy levels occurs when a two-level system is driven through an avoided level crossing \cite{lan32,zen32,iva23}. In that case, the LZ formula gives the transition probability between the instantaneous energy eigenstates or the adiabatic states, 
\begin{equation}
P_{{\rm LZ}}=\exp\left(-\pi\frac{\Omega^2}{2v}\right),
\label{lze}
\end{equation}
where $\Omega$ is the energy gap at the avoided crossing, and $v$ is the rate at which the avoided crossing is passed. For a slow quench ($v\to 0$), the population transfer is minimal ($P_{{\rm LZ}}\to 0$), while for a sudden sweep ($v\to\infty$), a complete transition ($P_{{\rm LZ}}\to 1$) occurs. The LZ transition according to Eq.~(\ref{lze}) is verified using Rydberg atoms \cite{pil84, noe98, rub81, con02, saq10, mae11, zha18}. An interacting pair of Rydberg atoms can emulate different LZ models \cite{bas18, ank20} and is relevant in implementing quantum gates \cite{hua17, hua18, jin21} and phenomena like population trapping \cite{mal21}. Crucially, LZ transitions can also generate entanglement \cite{qui13, tia15, qia17, bet20}. 

%%%%%%%%%%%%%%%%%%%%%%%%%
% FIGURE 1
%%%%%%%%%%%%%%%%%%%%
%
\begin{figure}
\centering
\includegraphics[width= .9\columnwidth]{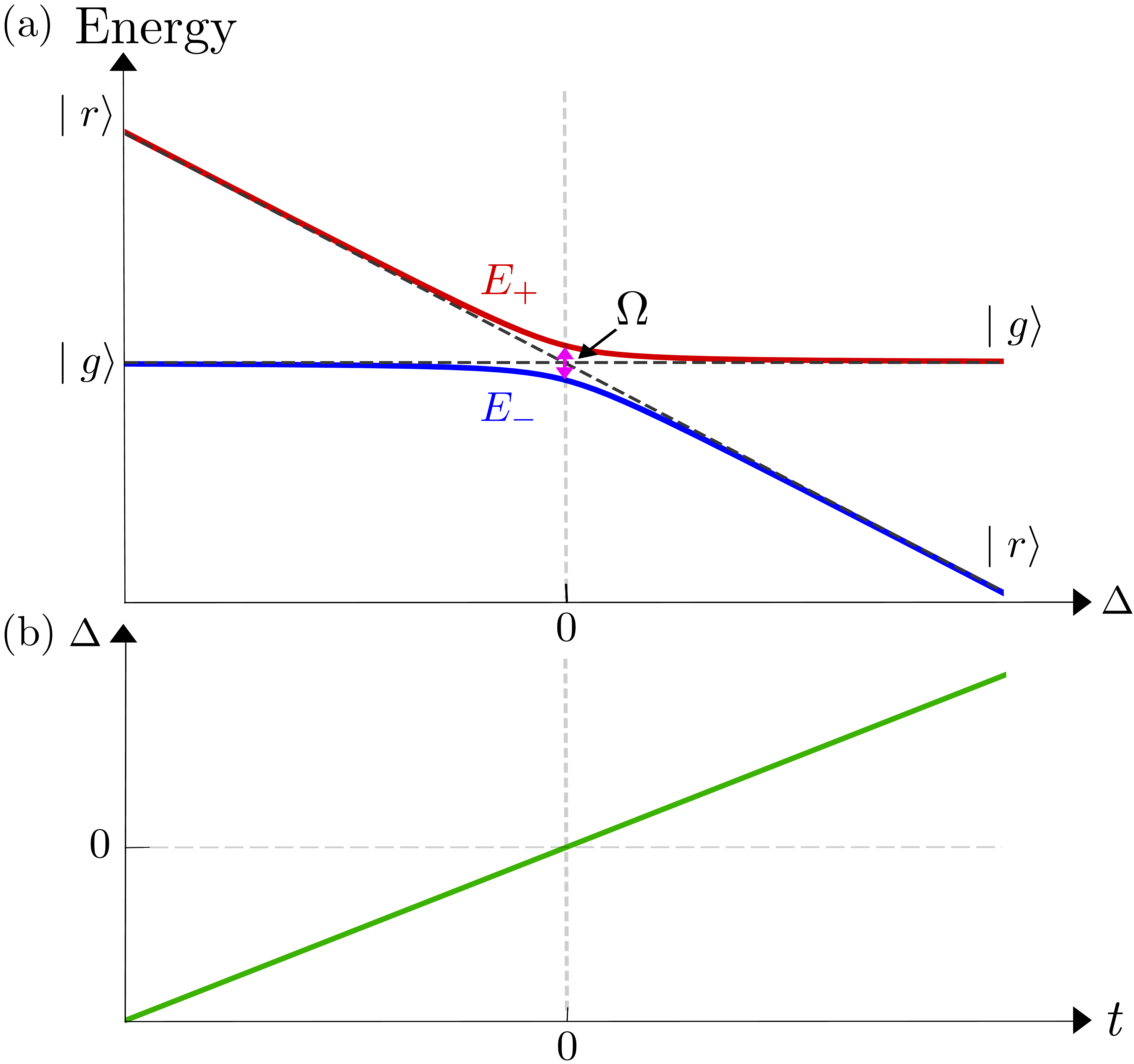}
\caption{\small{(a) Instantaneous energy eigenvalues resulting from linear variation of detuning in (b). The dashed lines in (a) show the diabatic energy levels. Far from the avoided level crossing, the diabatic and adiabatic levels merge. Across the avoided level crossing the LZ transition takes place.}}
\label{fig:1} 
\end{figure}
%%%%%%%%%%

%this paper
%
In this paper, we show that a maximally entangled Rydberg atom pair can be created by an LZ sweep from initially unentangled atoms for any non-zero RRIs. It is in stark contrast to the Rydberg blockade, where the maximally entangled Rydberg atom pair is created via strong RRIs. We characterize the entanglement between the atoms by the entanglement entropy for the coherent dynamics and quantum discord for the dissipative dynamics. Depending on the sweep rates and RRI strengths, various maximally entangled states are formed, and for a given set of parameters, the maximally entangled states change periodically. They are local unitary equivalent to the Bell states, and we explicitly verify that by calculating corresponding polynomial invariants for the two-qubit states \cite{lin99,mak02,jin15, cui17}. Finally, considering the spontaneous emission, we show that the maximally entangled states via LZ sweeps can be realized using high-level rubidium Rydberg states.

%paper structure
%
The paper is structured as follows. In Sec. \ref{salz}, we discuss the LZ dynamics in a single two-level atom. A pair of Rydberg atoms under LZ sweep is studied in Sec.~\ref{tad}. In particular, the polynomial local unitary invariants are provided in Sec.~\ref{pluv}. The coherent dynamics and generation of maximally entangled qubit states are discussed in Sec.~\ref{cd}. The effect of spontaneous emission from the Rydberg state is discussed in Sec.~\ref{dcd}. Finally, we summarize and provide an outlook in Sec.~\ref{co}.

%%%%%%%%%%%%%%%%%%%%%%%%%%%%%%%%%%%
\section{Single atom}
\label{salz}

% dynamics
First, we briefly discuss the LZ dynamics in a single two-level atom. The Hamiltonian describing a single two-level atom with a time-dependent detuning is
\begin{equation}
\hat H(t)=-\Delta(t)\hat\sigma_{rr}+\frac{\Omega}{2}\hat\sigma_x,
\label{ham1}
\end{equation}
where $\hat\sigma_{ab}=|a\rangle\langle b|$ with $a, b\in \{g, r\}$ includes both transition and projection operators, $\hat\sigma_x=\hat\sigma_{rg}+\hat\sigma_{gr}$, $\Omega$ is the constant Rabi frequency and $\Delta(t)=vt$ is the time dependent detuning with sweep rate $v$. The states $\{\ket{g}, \ket{r} \}$ form the diabatic basis, whereas the adiabatic basis consists of the instantaneous eigenstates of the Hamiltonian, $\hat{H}(t) \ket{\phi_\pm (t)} = E_\pm(t)\ket{\phi_\pm(t)}$. The time-dependent energy eigenvalues are $E_\pm (t) = \pm \frac{\Omega}{2}\beta_\mp(t)$ with $\beta_\pm (t) = \left[\bar{\Omega}(t)\pm \Delta(t)\right]/\Omega$ and $\bar{\Omega}(t) = \sqrt{\Delta(t)^2 + \Omega^2}$. The adiabatic and diabatic bases are related to each other by the time-dependent coefficients $\beta_\pm(t)$ via
\begin{equation}
\ket{\phi_\pm(t)} = \sqrt{\frac{\Omega}{2\bar{\Omega} }}\left( \pm \sqrt{\beta}_\pm \ket{g} + \sqrt{\beta}_\mp  \ket{r} \right).
\end{equation}
Far away from the avoided level crossings ($|\Delta|>>\Omega$), the adiabatic states coincide with the diabatic ones [see Fig. \ref{fig:1}(a)]. 

%%%%%%%%%%%%%%%%%%%%%%
%% FIGURE 2
%%%%%%%%%%%%%%%%%%%%%%
%
\begin{figure}
\centering
\includegraphics[width= 1.\columnwidth]{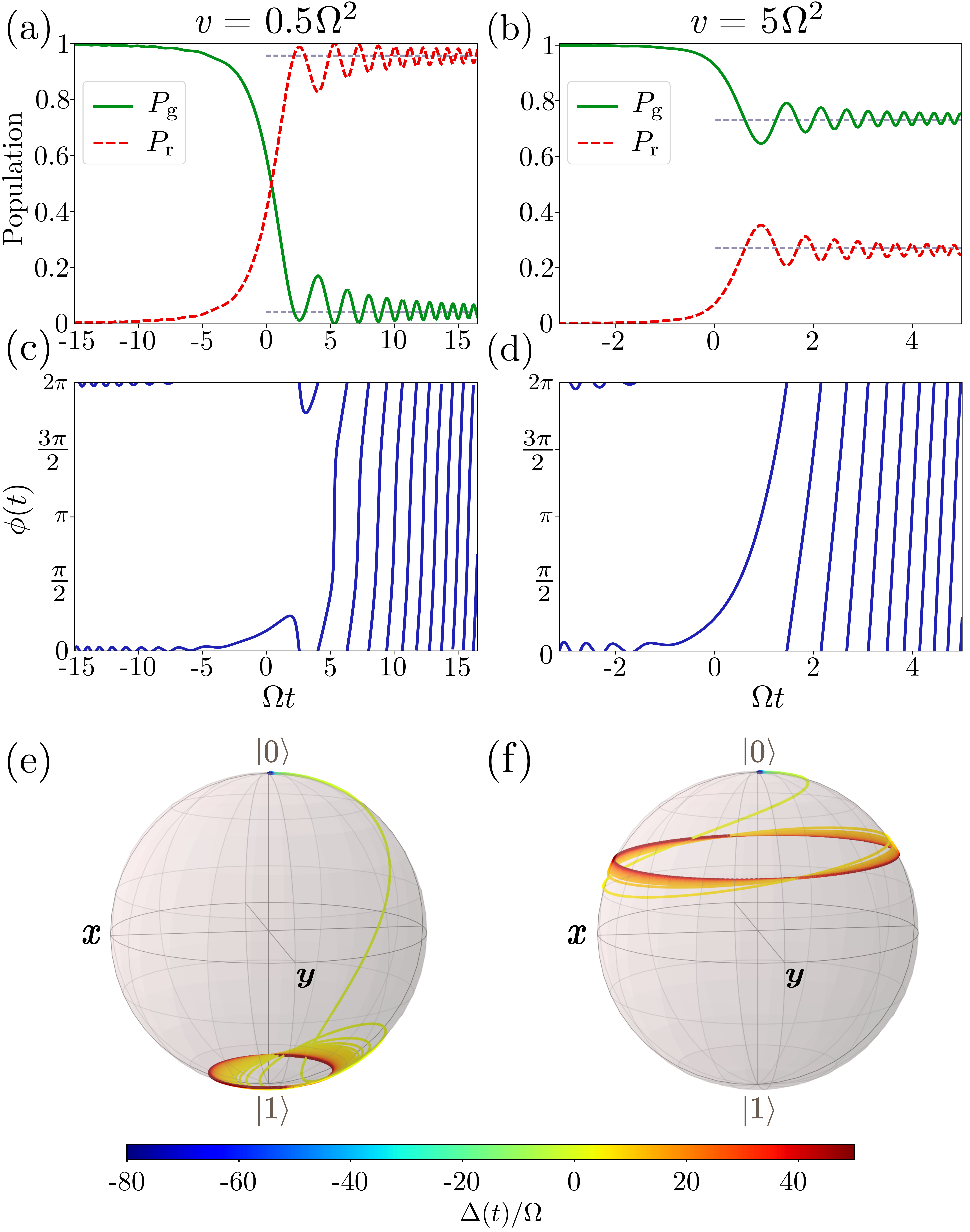}
\caption{\small{The population and phase dynamics in a two-level atom under an LZ sweep of the detuning for $v=0.5 \Omega^2$ and $v=5 \Omega^2$. (a) and (b) show the population dynamics where $P_{\rm g}=|\langle g|\psi(t)\rangle|^2=|a_g|^2$ and $P_{\rm e}=|\langle e|\psi(t)\rangle|^2=|a_e|^2$. (c) and (d) show the dynamics of the relative phase $\phi$. The trajectory of the Bloch vector on the unit sphere is shown in (e) and (f).}}
\label{fig:2} 
\end{figure}
%
% dynamics
Assuming the atom is initially in the lowest state $\ket{\phi_+}\sim |g\rangle$, the transition probability to the excited state $\ket{\phi_-(t)}$ after a sweep across the avoided level crossing at $\Delta=0$ is given by Eq.~(\ref{lze}). The exact dynamics of the system is obtained by numerically solving the Schr\"odinger equation: $i\partial/\partial t |\psi(t)\rangle=\hat H|\psi(t)\rangle$. For small enough sweeping rate, population from $|g\rangle$ adiabatically transfer to $\ket{r}$ [see Fig.~\ref{fig:2}(a)] and as $v$ increases the population transfer decreases [see Fig.~\ref{fig:2}(b)].  Equation~(\ref{lze}) agrees well with the exact results. Writing $|\psi(t)\rangle=a_{\rm g}(t)\ket{g}+a_{\rm r}(t)\exp({i\phi})\ket{r}$, the relative phase $\phi$ between $|g\rangle$ and $|r\rangle$ starts to vary as $\Delta(t)$ approaches the avoided crossing. After the LZ occurs, $\phi$ evolves continuously in time [see Figs.~\ref{fig:2}(c) and \ref{fig:2}(d)]. It indicates that even though the populations in $|g\rangle$ and $\ket{r}$ acquire a steady value as $t\to\infty$, the quantum state evolves through the relative phase $\phi$. The time evolution of the quantum state is more apparent in the Bloch sphere representation shown in Figs.~\ref{fig:2}(e) and \ref{fig:2}(f). For small $v$, the Bloch vector moves from one hemisphere to the other and eventually undergoes precession around a particular state [see Fig.~\ref{fig:2}(e)]. In contrast, for sufficiently large $v$, it remains in the same hemisphere [see Fig.~\ref{fig:2}(f)]. As we discuss below, for two qubits, the time dependence of the relative phases has far-reaching consequences.

%
%%%%%%%%%%%%%%%%%%%
%%%Two atom dynamics
%%%%%%%%%%%%%%%%%%%
%
\section{Two atoms}
\label{tad}
\begin{figure}
\centering
\includegraphics[width= 1.\columnwidth]{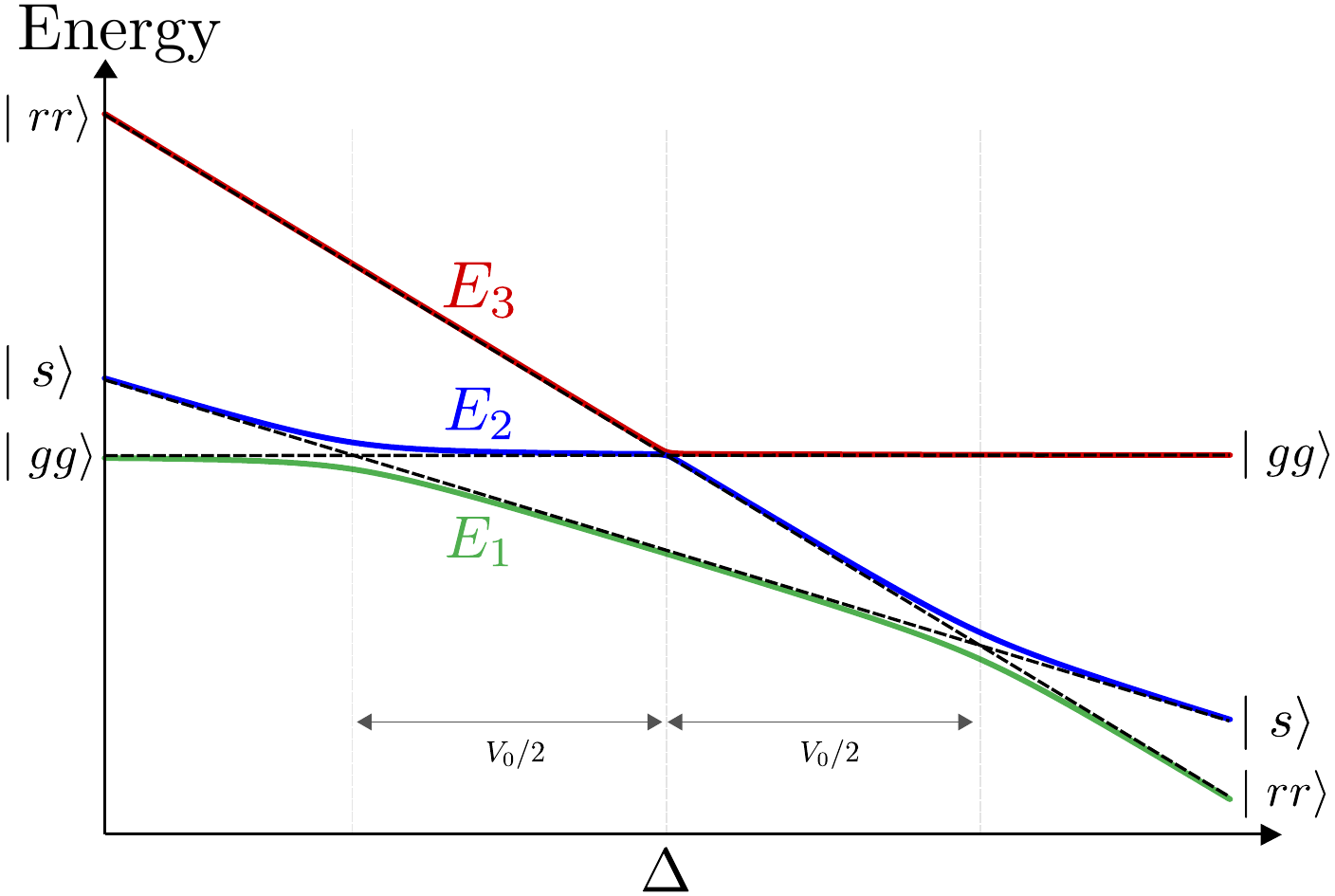}
\caption{\small{The energy eigenvalues of the two-atom Hamiltonian Eq.~(\ref{ham2}) for sufficiently large $V_0$ as a function of the detuning $\Delta$. The dashed lines show the diabatic energy levels and solid lines show the adiabatic energy levels. Far away from the avoided crossings the diabatic and adiabatic energy levels merge. }}
\label{fig:3} 
\end{figure}

The Hamiltonian of a pair of two Rydberg atoms is 
\begin{equation}
\hat H(t)=-\hbar\Delta(t)\sum_{i=1}^2\hat\sigma_{rr}^{i}+\frac{\hbar\Omega}{2}\sum_{i=1}^2\hat\sigma_x^{i}+V_0\hat\sigma_{rr}^{1}\hat\sigma_{rr}^{2},
\label{ham2}
\end{equation}
where $V_0$ is the RRI strength. The diabatic states of two-atoms are $\{\ket{gg}, \ket{s}, \ket{rr}\}$, where $\ket{s}=(|gr\rangle+|rg\rangle)/\sqrt{2}$. The system possesses three avoided crossings as shown in Fig.~\ref{fig:3} as function of $\Delta$, separated by $V_0/2$ \cite{bas18,ank20}. Up to a global phase factor, we can write the general state as 
\begin{equation}
|\psi(t)\rangle=a_{{\rm gg}}|gg\rangle+a_{{\rm s}}\exp({i\theta_1})|s\rangle+a_{{\rm rr}}\exp({i\theta_2})|rr\rangle, 
\label{g2s}
\end{equation}
where $\theta_{1, 2}$ are the time-dependent relative phases. We quantify the entanglement between the two atoms using the bipartite entanglement entropy $\mathcal S_{\rm A}(t)=-{\rm Tr}[\hat\rho_A(t)\log_2 \hat\rho_A(t)]=-\sum_{i=1}^2\lambda_i(t)\log_2\lambda_i(t)$ where $\hat\rho_A$ is the reduced density matrix of the first atom with its eigenvalues $\lambda_1=(1-x)/2$ and  $\lambda_2=(1+x)/2$, where $x=\sqrt{A+B\cos(2\theta_1-\theta_2)}$ with $A=(a_{{\rm gg}}^2-a_{{\rm rr}}^2)^2+2a_{{\rm s}}^2(a_{{\rm gg}}^2+a_{{\rm rr}}^2) $ and $B=4a_{{\rm gg}}a_{{\rm s}}^2a_{{\rm rr}}$. The eigenvalues $\lambda_i$ and hence $\mathcal S_{\rm A}$, depend only on the angle $2\theta_1-\theta_2$ when the coefficients $a_{{\rm gg}}, a_{\rm s}$ and $a_{{\rm rr}}$ are fixed. In that case, the entanglement between the atoms in the state $|\psi(t)\rangle$ is maximized when $2\theta_1-\theta_2=\pm(2n+1)\pi$, where $n=0, 1, 2, ...$. The maximum value depends on $V_0$ and $v$.

%%%%%%%%%%%%%%%%%%%%%%
%% FIGURE 3 Level scheme
%%%%%%%%%%%%%%%%%%%%%%
%
%

%%%%%%%%%%%%%%%%%%
%%% local invariants 
%%%%%%%%%%%%%%%%%%

\subsection{Polynomial local unitary invariants}
\label{pluv}
%%%%%%%%%%%%%%%%
The unitary equivalence between two-qubit states demands a set of twelve polynomial local unitary invariants \cite{lin99,mak02,jin15} and the determinant of a matrix $T_{12}$ defined below \cite{cui17} to be the same. Writing the two-qubit density matrix as 
\begin{eqnarray}
\rho=\dfrac{1}{2^2}I^{\otimes 2}+\sum_{j=1}^{2}\sum_{\alpha=1}^{3}T_{j}^\alpha\sigma_\alpha^{(j)}+\sum_{\alpha_1,\alpha_2=1}^{3}T_{12}^{\alpha_1\alpha_2}\sigma_{\alpha_1}^{(1)}\sigma_{\alpha_2}^{(2)},
\label{2rho}
\end{eqnarray}
where $\sigma_{1, 2, 3}$ are the Pauli matrices, $I$ is the identity matrix, $\sigma_\alpha^{(1)}=\sigma_\alpha\otimes I$, $\sigma_\alpha^{(2)}=I\otimes\sigma_\alpha$, $T_j$ is a three-dimensional real vector and $T_{12}^{\alpha_1\alpha_2}=\dfrac{1}{2^2}{\rm Tr}(\rho\sigma_{\alpha_1}^{(1)}\sigma_{\alpha_2}^{(2)})$ forms $3\times 3$ matrices. Two states $\rho$ and $\rho'$ are local unitary equivalent if and only if there are ${\rm SO(3)}$ operators $O_1$ and $O_2$ such that 
\begin{eqnarray}
T_1'=O_1T_1,\hspace{0.4cm} T_2'=O_2T_2\\
T_{12}'=(O_1\otimes O_2)T_{12}=O_1T_{12}O_2^{\rm T},
\end{eqnarray}
where T denotes the transpose of a matrix. It has been shown that two two-qubit states are local unitary equivalent if and only if they have the same values for the following 13 invariants \cite{jin15}: the inner products $\langle T_1,(T_{12}T_{12}^{\rm T})^\beta T_1 \rangle$, $\langle T_2,(T_{12}^{\rm T}T_{12})^\beta T_2 \rangle$, $\langle T_1,(T_{12}T_{12}^{\rm T})^\beta T_{12}T_2\rangle$, ${\rm Tr}(T_{12}T_{12}^{\rm T})^\alpha$ and ${\rm det}~T_{12}$ with $\beta=0, 1, 2$ and $\alpha=1, 2, 3$. We are interested in the maximally entangled states of a pair of Rydberg atoms formed by LZ sweeps and verifying their unitary equivalence to the Bell states.
%%%%%%%%%%%%%%%%%%%%%%%%%%%%%%%%%%%%%%
%%%dynamics
%%%%
\subsection{Coherent dynamics}
\label{cd}
A detailed analysis of the population dynamics subjected to LZ sweeps in a pair of Rydberg atoms can be found in \cite{ank20}. Here, we explore the phase and correlation dynamics of the system initially prepared in $|gg\rangle$ after passing through all three avoided level crossings. The LZ sweep builds quantum correlations between the atoms that are initially uncorrelated. For adiabatic evolution ($v\to 0$), the system eventually arrives at the product state $|rr\rangle$. Thus, the correlations which are built across the avoided crossings are lost at longer times. In the limit, $v\to \infty$, the system remains in $|gg\rangle$ and is again uncorrelated at longer times. Interesting scenarios emerge for intermediate values of $v$ where $\mathcal S_{\rm A}$ oscillates periodically.

\begin{figure}
\centering
\includegraphics[width= 1.\columnwidth]{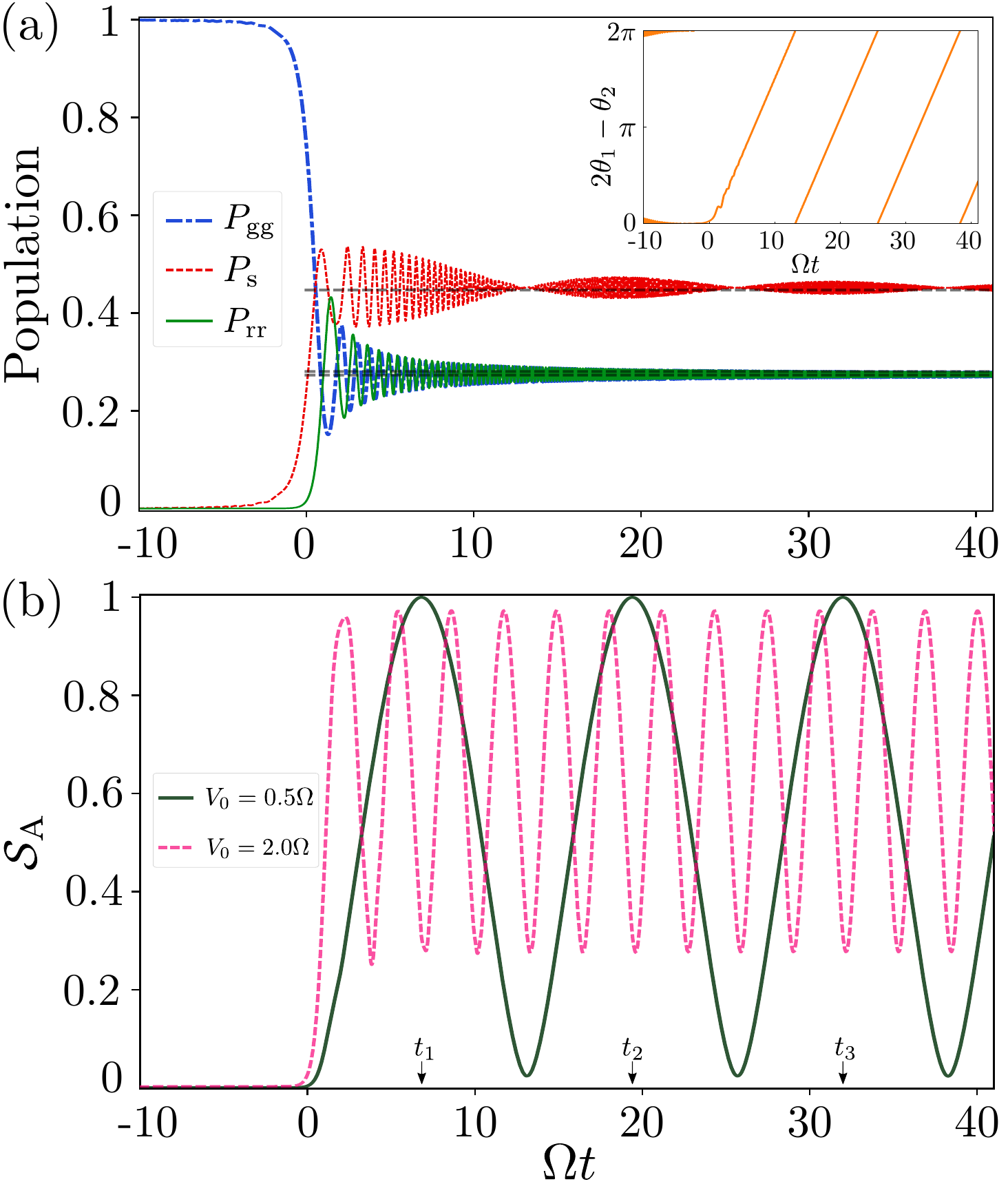}
\caption{\small{LZ dynamics in a pair of Rydberg atoms for weak RRIs,  $v=2.42\Omega^2$ and $\Delta(t_i)=100 \Omega$. (a) The dynamics of population in the diabatic states for $V_0=0.5\Omega$ and the solid line in (b) shows the bipartite entanglement entropy during the same dynamics as in (a). The horizontal dashed lines in (a) show the analytical results and the inset shows the evolution of the angle $2\theta_1-\theta_2$. The three instants at which $\mathcal S_{\rm A}\sim 1$ for $V_0=0.5\Omega$ are marked by $t_{1, 2, 3}$. Increasing to $V_0=2 \Omega$ (dashed line), the frequency of oscillation of $\mathcal S_{\rm A}$ is augmented by four times indicating its linear dependence on $V_0$.}}
\label{fig:4} 
\end{figure}

\begin{figure}
\centering
\includegraphics[width= .9\columnwidth]{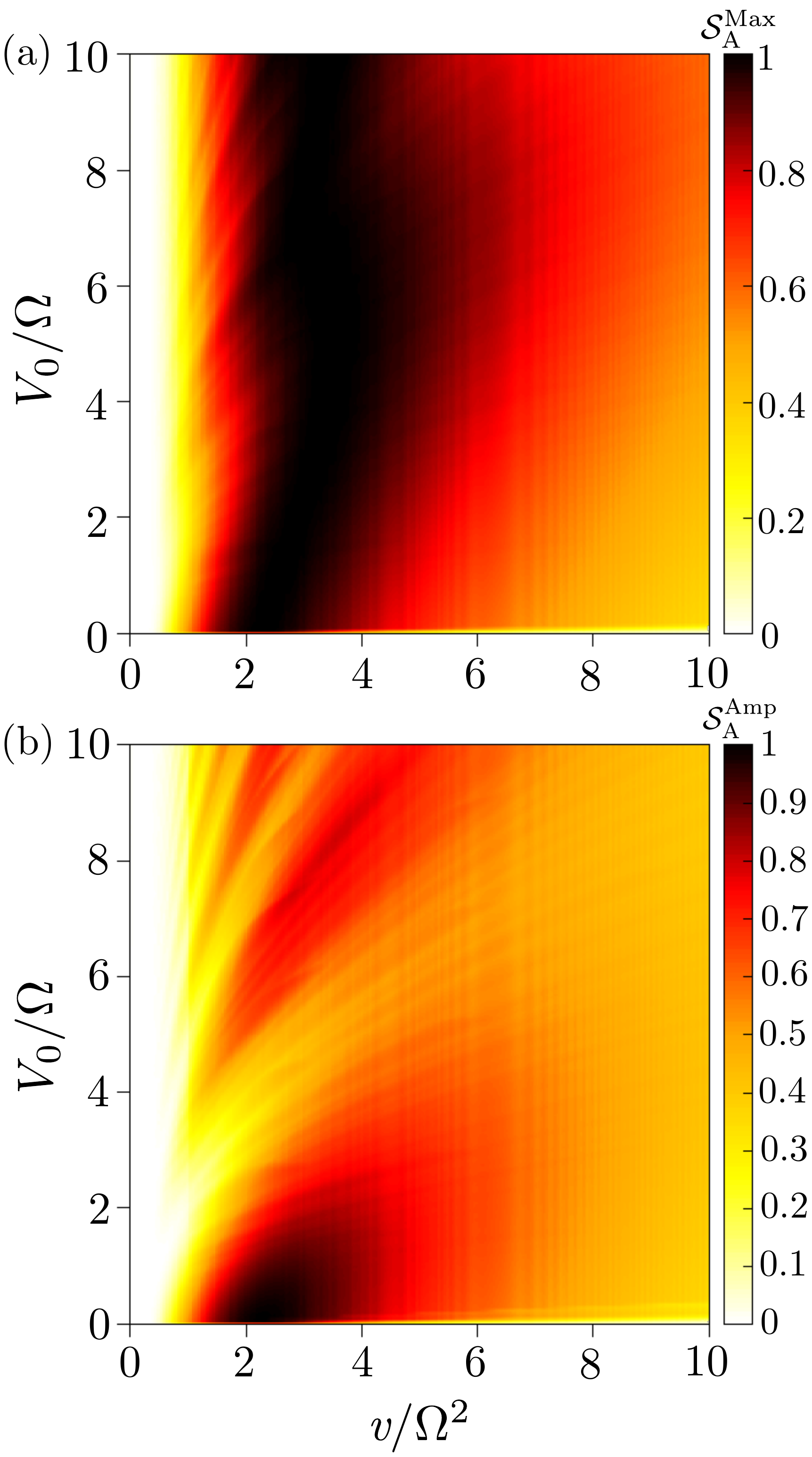}
\caption{\small{$\mathcal S_{\rm A}^{{\rm Max}}$ and $\mathcal S_{\rm A}^{{\rm Amp}}$ as a function of $V_0$ and $v$. The darker region in (a) indicates the emergence of maximally entangled state and that of (b) where we see a complete oscillation of $\mathcal S_{\rm A}$ between 0 and 1.}}
\label{fig:5} 
\end{figure}

%oscillatory entanglement dynamics
For sufficiently small RRIs, after passing through all three avoided crossings, the populations are given by $P_{{\rm gg}}(t\to\infty)=P_{{\rm LZ}}^2$ and  $P_{{\rm rr}}(t\to\infty)=1-Q_{{\rm LZ}}^2$, where $Q_{{\rm LZ}}=P_{{\rm LZ}}\exp(-\pi\Omega^2V_0/4v^{3/2})$ \cite{ank20}, shown by dashed horizontal lines in Fig.~\ref{fig:4}(a). Even though the populations eventually attain a steady value, the phase $2\theta_1-\theta_2$ evolves continuously over time as shown in the inset of Fig.~\ref{fig:4}(a). The latter leads to periodic oscillations in $\mathcal S_{\rm A}$ [see Fig.~\ref{fig:4}(b)]. Interestingly, for the chosen parameters ($V_0=0.5\Omega$ and $v=2.42\Omega^2$), $\mathcal S_{\rm A}$ attains a maximum value $\mathcal S_{\rm A}^{{\rm Max}}\sim 1$, i.e., the Rydberg atoms are maximally entangled and a minimum value $\sim 0$. Thus, the two atoms periodically entangle and disentangle over time. We observe that the oscillation frequency of $\mathcal S_{\rm A}$ depends (linearly) only on $V_0$ [see Fig.~\ref{fig:4}(b)], whereas the amplitude $\mathcal S_{\rm A}^{{\rm Amp}}$ of oscillations depends non-trivially on $V_0$ and $v$. In Fig.~\ref{fig:5}, we show $\mathcal S_{\rm A}^{{\rm Max}}$ and $\mathcal S_{\rm A}^{{\rm Amp}}$ which capture the dynamics of $\mathcal S_{\rm A}(t)$ entirely. Strikingly, as seen in Fig.~\ref{fig:5}(a), a maximally entangled state is possible to realize for any non-zero $V_0$ with an intermediate range of $v$. Thus, the LZ sweep can create maximally entangled states even for relatively weak RRIs, unlike the Rydberg blockade. It is also evident from Fig.~\ref{fig:5} that we can independently tune the amplitude and maximum of $\mathcal S_{\rm A}(t)$ by varying either $V_0$ or $v$. For $V_0^2/v<<1$, one can show that $\mathcal S_{\rm A}$ periodically becomes maximum at  $t_{n+1}\simeq[(2\theta_1-\theta_2)\big |_{t=V_0/v}+(2n+1)\pi]/V_0+V_0/v$, where the first term is computed at the third avoided crossing, which occurs at $t=V_0/v$.
%%%%%%%%%%%%%%%%%%%%%%
%% FIGURE 
%%%%%%%%%%%%%%%%%%%%%%
%

%%%%%%%%%%%%%%%%%%%%%%
%% FIGURE 4
%%%%%%%%%%%%%%%%%%%%%%
%
%
% entangled states to Bell states
%
Interestingly, every time the system periodically attains $\mathcal S_{\rm A}^{{\rm Max}}= 1$ as in Fig.~\ref{fig:4}(b) for $V_0=0.5 \Omega$, it is a different quantum state. The maximally entangled states seen at the three instants in Fig.~\ref{fig:4}(b) for $V_0=0.5\Omega$ are
\begin{eqnarray}
\label{mes1}
|\psi(t_1)\rangle= \begin{bmatrix} 0.54408 \\ 0.64031\exp(i*3.94928) \\ 0.54219\exp(i*4.75697)\end{bmatrix} \\
\label{mes2}
|\psi(t_2)\rangle= \begin{bmatrix} 0.52322 \\ 0.6748\exp(i*1.40484) \\ 0.52046\exp(i*5.95035)\end{bmatrix} \\
|\psi(t_3)\rangle= \begin{bmatrix} 0.530073 \\ 0.66414\exp(i*3.76878) \\ 0.52719\exp(i*4.39459)\end{bmatrix}.
\label{mes3}
\end{eqnarray}
Calculating the 13 invariants for the above states, only four are non-zero. In Table~(\ref{inv}), we show those for the above maximally entangled states, demonstrating that they are identical to those of Bell states. This confirms the local unitary equivalence between the Bell states and the maximally entangled states in Eqs.~(\ref{mes1})-(\ref{mes3}). 

\begin{equation}
\begin{tabular}{ | m{3.55 em} | m{1.5 cm}| m{1.6 cm} | m{1.85 cm}| m{1.65cm} | } 
  \hline
  Quantum state & ${\rm Tr}(T_{12}T_{12}^{\rm T})$ & ${\rm Tr}(T_{12}T_{12}^{\rm T})^2$ & ${\rm Tr}(T_{12}T_{12}^{\rm T})^3$ & ${\rm det}~T_{12}$ \\ 
  \hline
 $\psi(t_1)$ & 0.187499 &0.0117186 & 0.000732411 & -0.0156249 \\ 
  \hline
  $\psi(t_2)$ & 0.187498 & 0.0117185 & 0.000732399 & -0.0156248 \\ 
  \hline
   $\psi(t_3)$ & 0.187498 & 0.0117185 & 0.000732397 & -0.0156247 \\ 
  \hline
  Bell states & 0.1875 & 0.0117188 & 0.000732422 & -0.015625 \\ 
  \hline
\end{tabular}
\label{inv}
\end{equation}

%
%%%%%%%%%%%%%%%%%%%%%%%%%%%%%%%%%%%%%%%%%%
%% Dissipative dynamics
%%%%%%%%%%%%%%%%%%%%%%%%%%%%%%%%%%%%%%%%
%
%%%%%%%%%
\subsection{Dissipative dynamics}
\label{dcd}
To investigate the effect of spontaneous emission from the Rydberg state on the population and correlation dynamics, we solve the master equation for the two atom density matrix \cite{hai14}, 
\begin{equation}
\partial_t \hat{\rho} = -i \left[\hat{H}(t),\hat{\rho}\right]  +\mathcal{L} [\hat{\rho}],
\label{meq}
\end{equation}
with the Lindblad super operator given by 
\begin{equation}
\mathcal{L}[\hat \rho] =   \sum_{i=1}^2\hat C_i \hat{\rho} \hat C^{\dagger}_i- \frac{1}{2} \sum_i \left(\hat C^{\dagger}_i \hat C_i \hat{\rho} + \hat{\rho}\hat C^{\dagger}_i \hat C_i\right)
\end{equation}
where the decay operator, $\hat C_i= \sqrt{\Gamma} \hat\sigma_{ge}^i$ with $\Gamma$ as the spontaneous decay rate of the Rydberg state $|r\rangle$. The dissipative mechanism drives the system eventually into a mixed state. For a mixed state, the entanglement entropy $\mathcal S_{A}$ is no longer a good measure of quantum correlations since it fails to distinguish between classical and quantum correlations, and the quantum discord $\mathcal D(A:B)$ (see Appendix~\ref{qd}) is used \cite{sri19, kri21}. For $\Gamma=0$, we get $\mathcal D(A:B)=\mathcal S_{A}$.

%The population dynamics in the presence of spontaneous emission for different rubidium Rydberg states are discussed in Appendix~\ref{dpd}, and the dynamics of the discord is shown in Fig.~\ref{fig:6}.

%%%%%%%%%%%%%%%%%%%%%%
%% FIGURE 6
%%%%%%%%%%%%%%%%%%%%%%
%
\begin{figure*}
\centering
\includegraphics[width= 2.\columnwidth]{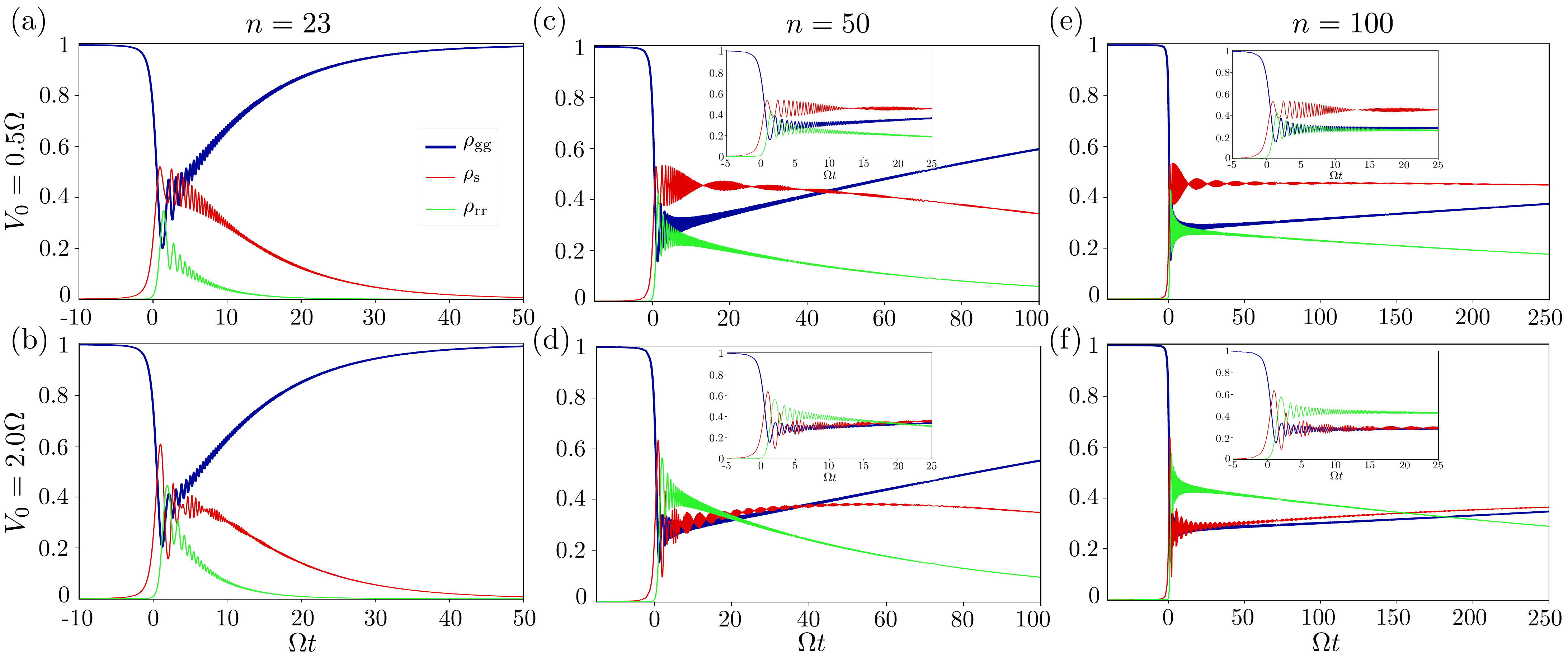}
\caption{\small{LZ dynamics in a pair of rubidium Rydberg states of $nS_{1/2}$ for $v=2.42\Omega^2$ and $\Delta(t_i)=100 \Omega$. Along each row, the principal quantum number $n$ changes and hence the decay rate $\Gamma$. Larger the $n$ smaller the decay rates $\Gamma$. The first row is for $V_0=0.5\Omega$ and second row is for $V_0=2\Omega$.  }}
\label{fig:6} 
\end{figure*}
%%%%%%%%%%%%%%

\begin{figure}
\centering
\includegraphics[width= .9\columnwidth]{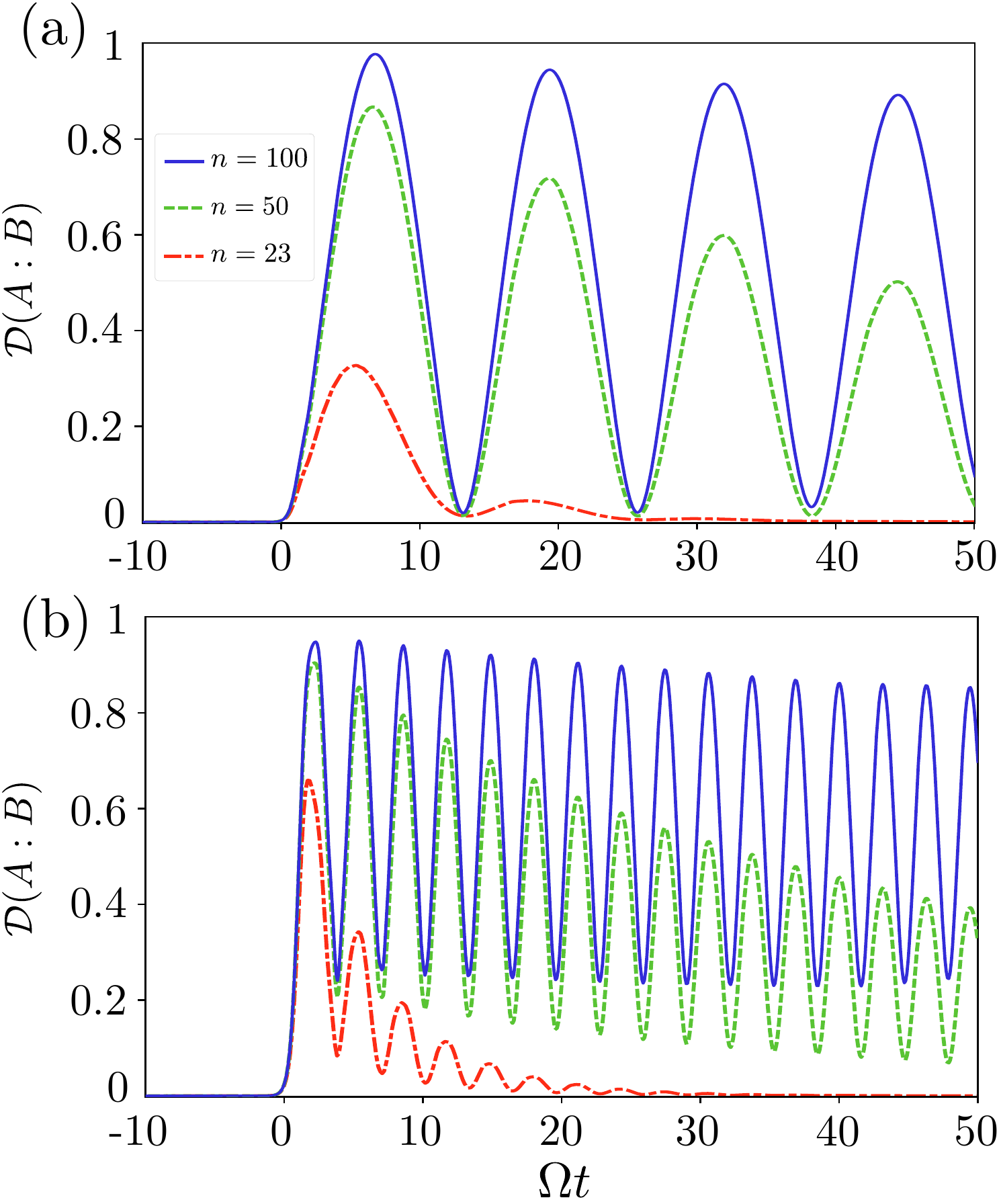}
\caption{\small{The dynamics of quantum discord $\mathcal D(A:B)$ for the same dynamics shown in Fig.~\ref{fig:6}. For (a) $V_0=0.5\Omega$ and for (b) $V_0=2\Omega$.}}
\label{fig:7} 
\end{figure}

The population dynamics for different rubidium Rydberg states is shown in Fig.~\ref{fig:6}. Note that the interaction strength $V_0$ can vary via inter-atom separation or $n$. For small $n$ [see Figs.~\ref{fig:6}(a) and \ref{fig:6}(d) for $n=23$], the decay rate is significant, and the system attains the steady state ($\rho_{{\rm gg}}=1$) relatively quickly. For larger $n$, the decay rate from the Rydberg state is low, and the reminiscence of coherent dynamics becomes evident at shorter times. 
Numerically, we see that $\rho_{{\rm gg}}(t\to\infty)\propto 1-\exp(-c_1\Gamma t)$, $\rho_{{\rm s}}(t\to\infty)\propto \exp(-c_2\Gamma t)$ and $\rho_{{\rm rr}}(t\to\infty)\propto \exp(-c_3\Gamma t)$, where the constants $c_{1, 2, 3}$ shows a $\sqrt{v}$ dependence for a fixed $V_0$ and varies linearly in $V_0$ with a negative slope for a given $v$. As expected, the quantum discord $\mathcal D(A:B)$, where $A$ and $B$ label the atoms, exhibits decaying oscillatory behaviour as shown in Fig.~\ref{fig:7}. The decay constant $\Gamma$ reduces the maximum of $\mathcal D(A:B)$ but leaves its oscillation frequency intact. Considering the spontaneous emission from the Rydberg state, generating maximally entangled states via LZ sweeps is preferable to having a high $n$ state.

%%%%%%%%%%%%%%%%%%%%%%
%% FIGURE 7
%%%%%%%%%%%%%%%%%%%%%%
%
%
%%%%%%%%%%%%%%%%%%%%
\section{Conclusion and outlook}
\label{co}
We analyzed the creation of maximally entangled states in a Rydberg atom pair through LZ sweeps starting from an initial product state $|gg\rangle$. Unlike the Rydberg Blockade, where strong RRIs are required, LZ sweeps generate a maximally entangled state even for low RRIs. Under an LZ sweep, an atom pair evolves periodically through various maximally entangled states. The local unitary equivalence between them and the Bell states is verified by evaluating the polynomial invariants. Incorporating the spontaneous emission, we show that high-lying rubidium Rydberg states are best suited. Note that, not only for Rydberg atoms, our results are equally valid for any coupled qubits, such as NMR qubits \cite{kri21}. Our studies can be extended beyond two atoms, and to many body systems to explore the creation of exotic, highly entangled quantum matter in Rydberg atom arrays \cite{bro20} via LZ sweeps.
%%%%%%%

\section{Acknowledgements}
We acknowledge useful discussions with Chirag Gupta, Ankita Niranjan, Yashwant Chougale, V R Krithika and T S Mahesh. We thank National Supercomputing Mission (NSM) for providing computing resources of "PARAM Brahma" at IISER Pune, which is implemented by C-DAC and supported by the Ministry of Electronics and Information Technology (MeitY) and Department of Science and Technology (DST), Government of India. R.N. further acknowledges DST-SERB for Swarnajayanti fellowship File No. SB/SJF/2020-21/19, the MATRICS grant (MTR/2022/000454) from SERB, Government of India and National Mission on Interdisciplinary Cyber-Physical Systems (NM-ICPS) of the Department of Science and Tech- nology, Government of India, through the I-HUB Quantum Technology Foundation, Pune, India. D.V. thanks Department of Science and Technology (India) for INSPIRE fellowship. W. L. is funded by  the EPSRC (Grant No. EP/W015641/1). This work is partially funded by the Going Global  Partnerships Programme of the British Council (Contract No.IND/CONT/G/22-23/26). We further thank QUTIP python library for the numerical calculations of open quantum systems \cite{joh12} and open source library "ARC" \cite{sib17}.

%%%%%%%%%%%%%%%
%Appendix
%%%%%%%%%%%%%%%%
\appendix
%%%%%%%%%%%
\section{Quantum discord}
\label{qd}
To define the quantum discord, we briefly sketch the mutual information in classical information theory. The classical mutual information between two subsystems $A$ and $B$ is defined as $\mathcal I= H_A+H_B-H_{AB}$ where $H_A$ ($H_B$) is the Shannon entropy of the subsystem $A$ ($B$), and $H_{AB}$ is the joint entropy of $A$ and $B$. An equivalent expression for mutual information is $\mathcal J=H_B-H_{B|A}$, where $H_{B|A}$ is the conditional entropy, the information needed to describe $B$ when $A$ is known. While, in the classical theory $\mathcal I=\mathcal J$, in the quantum version, in which the Shannon entropy is replaced by the von Neumann entropy, there exists a discrepancy between $\mathcal I$ and $\mathcal J$ which is quantified by the quantum discord.

In the quantum theory, we have $\mathcal I= S_A+S_B-S_{AB}$ and $\mathcal J(B:A)=S_B-S_{B|A}$, where $\mathcal S_{AB}=-{\rm Tr}(\hat\rho\log_2\hat\rho)$ is the von Neumann entropy for the state $\hat\rho$, and $\mathcal S_{AB}=0$ for a pure state. Given a complete set of von Neuman projective measurements $\{\hat \Pi_A^i\}$ on the subsystem $A$ with probabilities $\{p^i\}$, the conditional entropy of the subsystem $B$ is $\mathcal S_{B|A}=\sum_ip^i\mathcal S_{B|i}$, where $\mathcal S_{B|i}$ is the von Neumann entropy for the reduced density operator $\hat\rho_{B}^i= {\rm Tr}_{A} \left[(\hat\Pi_{A}^i\otimes\mathbb I_B)\hat\rho_{AB}(\hat\Pi_{A}^i\otimes\mathbb I_B)^{\dagger}\right]/p^i$ with $p^i= {\rm Tr}_{AB} \left[(\hat\Pi_{A}^i\otimes\mathbb I_B)\hat\rho_{AB}(\hat\Pi_{A}^i\otimes\mathbb I_B)^{\dagger}\right]$ and $\mathbb I_B$ is the identity operator. It has been shown that the total classical correlation can be obtained as, $\tilde{\mathcal J}(B:A)=\max_{\{\hat\Pi_A^i\}}\left[S_B-\sum_ip^i\mathcal S_{B|i}\right]$. The maximization ($\max_{\{\hat\Pi_A^i\}}$) is carried across all the possible orthonormal measurement bases $\{\hat\Pi_A^i\}$ of the subsystem $A$. Similarly one can obtain $\tilde{\mathcal J}(A:B)$  where the measurements are being carried out on the subsystem $B$. Finally, the quantum discord is defined on both ways by swapping $A$ and $B$ as, 
\begin{equation}
 \mathcal D(A:B)=\mathcal I-\tilde{\mathcal J}(A:B), 
 \end{equation}
 and 
 \begin{equation}
 \mathcal D(B:A)=\mathcal I-\tilde{\mathcal J}(B:A).
 \end{equation}
 Note that, the quantum conditional entropy depends on the choice of the observables being measured on the other subsystem, and this results in a discrepancy between $\mathcal I$ and $\mathcal J(B:A)$ or $\mathcal J(A:B)$, which is quantified as the quantum discord. For a bipartite pure state $|\psi(t)\rangle$ the quantum discord coincides with the entanglement entropy, i.e.,  $\mathcal D(A:B)=\mathcal D(B:A)=\mathcal S_A=\mathcal S_B$.

%%%%%%%%%%%%%%%%%%%%%%%%

%%%%%%
\bibliographystyle{apsrev4-1}
\bibliography{lib.bib}
\end{document}